\input harvmac.tex

\lref\rWDB{E. Witten,  Nucl. Phys. {\bf B460} (1996) 335.}
\lref\rWSYMM{E. Witten,  Nucl. Phys. {\bf B202} (1982) 253.}
\lref\rSMP{A. Sen, Int. J. Mod. Phys. {\bf A9} (1994) 3707.}
\lref\rDH{A. Dabholkar and J. Harvey, Phys. Rev. Lett. {\bf 63} (1989) 719.}
\lref\rGH{J. Gauntlett and J. Harvey, ``S-Duality and the Spectrum of Magnetic
Monopoles in Heterotic String Theory", hep-th/9407111.}
\lref\rWSI{E. Witten, Nucl. Phys. {\bf B460} (1996) 541.}
\lref\rSM{B. Simon, Ann. Phys. {\bf 146} (1983), 209.}
\lref\rMP{M. Porrati, ``How to Find H-monopoles in Brane Dynamics'',
hep-th/9607082.}
\lref\rCallias{C. Callias, Commun. Math. Phys. {\bf 62} (1978), 213.}

\overfullrule=0pt

\def\a{{\alpha}}
\def\b{{\beta}}

\def\A{{Q_+}}
\def\B{{Q_-}}
\def\t{{\, t \,}}
\def\th{{\theta}}

\def\frac#1#2{{#1\over #2}}

%
%
\def\eqnn#1{\xdef #1{(\secsym\the\meqno)}\writedef{#1\leftbracket#1}%
\global\advance\meqno by1\wrlabeL#1}
\def\eqna#1{\xdef #1##1{\hbox{$(\secsym\the\meqno##1)$}}
\writedef{#1\numbersign1\leftbracket#1{\numbersign1}}%
\global\advance\meqno by1\wrlabeL{#1$\{\}$}}
\def\eqn#1#2{\xdef #1{(\secsym\the\meqno)}\writedef{#1\leftbracket#1}%
\global\advance\meqno by1$$#2\eqno#1\eqlabeL#1$$}

\Title{hep-th/9607145, HUTP-96/A015, DUK-M-96-3}{\vbox{\centerline{A Comment on
the Spectrum of H-Monopoles}}}
\centerline{S. Sethi\footnote{$^\ast$} {Supported in part by the Fannie and
John
Hertz Foundation (sethi@string.harvard.edu).} }
\medskip\centerline{\it Lyman Laboratory of Physics}
\centerline{\it Harvard University}\centerline{\it Cambridge, MA 02138, USA}
\vskip 0.15in
\centerline{and}
\vskip 0.15in
\centerline{M. Stern\footnote{$^\dagger$} {Supported by NSF Grant DMS 9505040
(stern@math.duke.edu).} }
\medskip\centerline{\it Department of Mathematics}
\centerline{\it Duke University}\centerline{\it Durham, NC 27706, USA}

\vskip .3in

We consider the question of  ground states in the supersymmetric system that
arises in the search for the missing H-monopole states.   By studying the
effective theory near certain singularities in the five-brane moduli space, we
find the remaining BPS states required by the conjectured S-duality of the
toroidally compactified heterotic string.

\Date{7/96}
\newsec{Introduction}

The toroidally compactified heterotic string provides a natural string theory
in which to conjecture the existence of an exact strong-weak coupling duality.
Some of the testable consequences of such a duality have been described in
detail by Sen \rSMP. Indeed, substantial evidence has accumulated for S-duality
in the field theoretic limit; yet, attempts at a truly stringy verification of
S-duality have created some puzzles.  We shall consider the heterotic theory at
a general point in its moduli space, where the gauge group is abelian. The
electrically charged states of interest are constructed by tensoring the right
moving ground state with an arbitrary left moving state. To saturate the BPS
bound, the constraint,
$$ N_L - 1 = {1\over 2} (p_R^2 - p_L^2), $$
must be satisfied \rDH.   Our interest resides in the twenty-four predicted
H-monopole states with $ N_L=1$. The number twenty-four just corresponding to
the choice of left moving oscillator. The first study of the H-monopole
spectrum, performed in \rGH, encountered problems reconciling restrictions on
the allowed structure of the H-monopole moduli space with the required number
of normalizable modes. Recently, Witten shed a great deal of light on the
question of the H-monopole spectrum in his study of small instantons \rWSI. By
a careful study of supersymmetric ground states in five-brane quantization at
generic points in the moduli space, Witten finds eight of the desired
twenty-four states. The purpose of this letter is to show that the remaining
sixteen states arise from the singularities in the five-brane moduli space
where the $ SU(2)$ gauge symmetry, arising from small instantons,  is still
broken to $ U(1)$, but a charged hypermultiplet becomes light, as conjectured
in \rWSI.  There are sixteen such singularities, so we desire a single
normalizable ground state in the effective theory near the singularity.

The problem involves a potential with flat directions extending to infinity
which somewhat complicates the analysis. There is a further issue of gauge
invariance since, near the singularity, we must include the charged
hypermultiplet in our analysis.  The particular system that arises in studying
the question of the missing H-monopole states is a special case of a more
general class of theories. The Hamiltonian for these systems takes the general
form,
\eqn\hamilt{H = {1\over 2} \Tr (p^i p^i) + V(x) + H_F.}
The bosonic potential $ V(x)$ is polynomial in $ x$, and generally has flat
directions. The term $ H_F$ is quadratic in the fermions and linear in $ x$.
The coordinates $ x$ are charged under the gauge group, which is generally
non-abelian, and the trace is over the gauge indices. Models of this kind
appear, for instance, in the study of D-brane bound states \rWDB.
We shall not undertake an analysis of the most general case. Such an analysis
is very interesting, but quite subtle. Rather, we shall use certain features of
this particular model that simplify the counting of the number of normalizable
ground states for the effective theory near the singularity.

Consequently, we shall restrict to the the case where the gauge group is
abelian. Since there are charged fields in the model under consideration, the
supersymmetry algebra no longer closes on the Hamiltonian.  Rather, the
supersymmetry algebra closes if a constraint, $ C = C^b + C^f$, following from
Gauss' law, is set to zero.  The gauge constraint, $ C$, splits into two $
U(1)$ generators: one generates rotations of the charged bosons, $ C_b$,  while
the other generates rotations of the charged fermions, $ C_f $.  As usual, we
are interested in counting the number of $ L^2$ ground states weighted by $
(-1)^F$ where $ F$ is the fermion number. Therefore, we wish to compute the
index,
\eqn\inddef{\eqalign{
 {\rm Ind}&= \int{ dx \, \lim_{\b \rightarrow  \infty} \, \tr \, (-1)^F e^{-\b
H} (x,x),} \cr
 &=n_B - n_F,\cr
}}
where the trace is over the gauge invariant spectrum of the Hamiltonian i.e.
states $ |\psi (x)>$ satisfying $C |\psi (x)> = 0$. Since the space of scalars
is non-compact and the potential has flat directions, the integral depends on $
\b$, and usually, we cannot just consider the more easily evaluated $ \b
\rightarrow 0$ limit.
Our first task is to implement the projection onto gauge invariant states
explicitly, so we can trace over the full, unconstrained spectrum. Let us
denote the operator generating a gauge transformation $ g(\theta ) $ on the
fermions by $ \Pi (g(\theta ))$ where we shall drop the explicit dependence on
$ \theta$. To project onto gauge invariant states, we insert:
\eqn\projection{\eqalign{
 {\rm I}( \b ) &=\int_{U(1)} d\theta {\int{ dx \, \tr \,  e^{i \theta C} \,
(-1)^F e^{-\b H} (x,x),}} \cr
 &= \int_{U(1)} d\theta {\int{ dx \, \tr \,  \Pi (g) \, (-1)^F e^{-\b H}
(gx,x)}} ,\cr
}}
where the measure for the $ U(1)$ integration is chosen so that $ \int_{U(1)}
d\theta = 1$. The trace is now over the full Hilbert space, including
gauge-variant states. Now we can choose a supersymmetry generator, $Q$, obeying
$ H=Q^2$ under the assumption $ C=0$. We can then write our index as,
\eqn\indbound{\eqalign{
{\rm Ind} &= \lim_{\b \rightarrow 0} \int dx  \int_{U(1)}  d\theta \, \tr \,
(-1)^F e^{-\b H} \Pi (g) \, (gx,x) \, + \cr
 & \phantom{=} \int dx \int_{U(1)} d\theta \int_0^\infty d\b \, \tr \,   (-1)^F
 Q^2 e^{-\b H} \Pi (g) \, (gx,x). \cr
}}

As we shall argue, the second term does not appear to contribute to the
index for this particular system.  The first term in \indbound\  can easily be
reduced to quadrature using perturbation theory, as we shall describe in the
following section.

\newsec{Counting H-monopole States}

The model of interest is the dimensional reduction of abelian supersymmetric
Yang-Mills from $5+1$ dimensions to $ 0+1$ with a single charged
hypermultiplet.  Rather than reduce from six dimensions, we shall, for
convenience,  reduce N=2 Yang-Mills from four dimensions. This choice hides the
symmetry between the scalars coming from the reduction of the six dimensional
connection, but it doesn't affect our analysis in any significant way.  From
reducing the (four-dimensional) vector multiplet and Higgs field, we obtain
three real scalars $ x^a$, and a complex scalar $ y$. The hypermultiplet
provides two more complex coordinates $ Q_+$ and $ Q_-$, where the subscripts
denote the $ U(1)$ charge. Introducing canonical momenta obeying,
$$ [ x, p_x ] = i, $$
we find that the Hamiltonian for this system takes the form:
\eqn\hamil{ \eqalign{ H = &  {1\over 2} p^a p^a + p_y p_y^\dagger + p_+
p_+^\dagger +  p_- p_-^\dagger
+  {1\over 2} (Q_+ Q_+^\dagger + Q_- Q_-^\dagger)^2 \cr & + (x^a x^a + 2 y
y^\dagger)  (Q_+ Q_+^\dagger + Q_- Q_-^\dagger) + H_F. } }
The term $ H_F$, which we shall describe below,  contains operators quadratic
in the fermions  and linear in the coordinates. The crucial point about the
bosonic potential is the existence of flat directions, occuring when $ Q_+ =
Q_- = 0$. Such flat directions complicate the counting of ground states since a
normalizable ground state decays with a power law along the flat directions,
rather than with the usual exponential fall-off.

The situation is actually somewhat more subtle than the preceding comment might
imply. If one were to consider just the bosonic theory, then the spectrum for
this model is actually discrete -- regardless of the flat directions in the
potential \rSM.  To construct a scattering state along the flat direction, we
would want to put the oscillators, which are transverse to the flat directions,
into their ground states; however, the zero point energy for these oscillators
increases as we travel down the flat direction, prohibiting finite energy
scattering states. Unfortunately, this argument is no longer true for the
supersymmetric theory, since the additional fermionic oscillators, required by
supersymmetry, cancel the bosonic zero point energy.

Let us introduce a set of two component fermions,  $ L, N, M_-, M_+$, each of
which obey an anti-commutation relation of the form,
$$ \{ L_\a, L_\b^\dagger \} = \delta_{\a\b}.$$
The fermionic term, $ H_F$, is given by,
\eqn\hferm{\eqalign{H_F = & \,  x^a ( M_+^\dagger \sigma^a M_+ - M_-^\dagger
\sigma^a M_-)
+ \sqrt{2} (y M_- \t M_+ -  y^\dagger M_-^\dagger \t M_+^\dagger ) \cr &
+  \sqrt{2} (\A M_+^\dagger \t L^\dagger -  \A^\dagger M_+ \t L-  \B
M_-^\dagger \t L^\dagger +  \B^\dagger M_-  \t L) \cr & +  \sqrt{2} ( \B N \t
M_+ + \A N \t M_-  -   \B^\dagger N^\dagger \t M_+^\dagger -  \A^\dagger
N^\dagger \t M_-^\dagger ) ,
} }
where $ \sigma^a$ are the Pauli matrices. The matrix  $ t = \pmatrix{
0 & -1 \cr
1 & 0 \cr
}, $ and summation is implied on all indices.

The component of the constraint which generates gauge transformations on the
charged bosons, $ C^b$,  is proportional to the operator,
$$ \B^\dagger \pi_-^\dagger - \A^\dagger \pi_+^\dagger + \A \pi_+ - \B \pi_- ,
$$
while the component generating transformations on  the fermions, $ C^f$, is
proportional to,
$$ M^\dagger_+ M_+ - M_-^\dagger M_-. $$
Our task is to show the existence of a normalizable state satisfying $ H |\psi
> = 0$, and $ C  |\psi > = 0$.

To count the number of ground states with sign, we begin by computing the
principal term in \indbound,
$$ {\rm I}( 0 )  = \lim_{\b \rightarrow 0} \int dx \, dy \, dQ
\int_{-\pi}^{\pi} {d\theta \over 2\pi}  \tr  (-1)^F e^{-\b H} \Pi (g) \,
(gQ,Q).
$$
The projection operator $ \Pi (g)$ is $ e^{i \th C^f}$.  To compute this term,
we shall approximate $ e^{-\b H}$ by the operator,
$$ {1\over (2\pi \b )^{9/2}} e^{ - (|x-x'|^2 +2 |y-y'|^2 +2 | e^{i\th}\A - {\A
}'|^2 + 2| e^{-i\th}\B - {\B }'|^2 ) / 2\b} \times $$ $$ e^{-\b V} e^{-\b H_F}
,  $$
where $ V$ is the bosonic potential given in \hamil. Corrections to this
approximation are suppressed by powers of $ \b $, and give a vanishing
contribution to the principal term. As usual, the inclusion of $ (-1)^F$ in the
trace creates fermion zero modes which must be absorbed to obtain a
non-vanishing contribution.  Fermions appear from two sources: $ H_F$, and the
constraint $ C^f$.  To obtain the required number of  $ L $ and $ N$ zero
modes, all fermion zero modes must be supplied by $ H_F$, rather than $ C^f$.
A non-vanishing contribution then arises when $ \Pi (g)$ is set to one.  The
integral now becomes,
$$   \lim_{\b \rightarrow 0} \int  \,  \, \int_{-\pi}^{\pi} {d\theta \over
2\pi} \, {1\over (2\pi \b )^{9/2}} e^{ - ( | e^{i\th}\A - {\A }|^2 + |
e^{-i\th}\B - {\B }|^2 ) / \b}  \times $$ $$ e^{-\b V} {1\over 8!}\tr \, (-1)^F
( \b H_F )^8,
$$
where the explicit trace is now only over the fermion modes.  By rescaling all
of the scalar coordinates, we can replace the integral by the expression,
$$ \lim_{\b \rightarrow 0} \int  \,  \, \int_{-\pi}^{\pi} {d\theta \over 2\pi}
\, \b^{15/4} \, {1\over (2\pi \b )^{9/2}} e^{ - ( | e^{i\th}\A - {\A }|^2 + |
e^{-i\th}\B - {\B }|^2 ) / \b^{3/2}}  \times $$ $$ e^{- V} {1\over 8!} \tr \,
(-1)^F ( H_F )^8.
$$
We can rewrite $ | e^{i\th}\A - {\A }|^2 $ as $ 2 | \A |^2 (1-  \cos{\th} )$,
which can further be replaced by $ | \A |^2 \th^2$ as $ \b $ becomes small. On
rescaling $ \th$, we are left with the integral,
$$  \int  \,  \, \int_{-\infty}^{\infty} {d\theta \over 2\pi} \,   {1\over
(2\pi )^{9/2}} e^{ -\th^2 ( | \A |^2 + | \B |^2 ) }  e^{- V} {1\over 8!} \tr \,
(-1)^F ( H_F )^8,
$$
which fortunately is $ \b $-independent. Note that without the projection onto
gauge-invariant states, we would not have arrived at an expression independent
of $ \b$. A preliminary investigation of  $ H_F$, shown in \hferm,  shows that
the terms proportional to $ x$ or $ y$ cannot contribute the required number of
fermions for a non-vanishing trace. The integrations over $ x, y, \th$ are then
Gaussian giving a total factor of,
$$ {1\over 2 (2 \pi )^{9/2}} { \pi^2 \over   ( | \A |^2 + | \B |^2 )^3} {1\over
8!}, $$
and leaving only the integration over $ \A$, and $ \B$.

We now require the term in $ ( H_F )^8$ proportional to the `volume form' for
the fermion zero modes.  A non-vanishing contribution comes from the following
terms in  $ ( H_F )^8$,
$$  16 \left (\matrix{
8\cr
4\cr
}\right ) \left (\matrix{
4\cr
2\cr
}\right )^2 ( \B^\dagger M_-  \t L -  \A^\dagger M_+ \t L )^2   (\A M_+^\dagger
\t L^\dagger -  \B M_-^\dagger \t L^\dagger  )^2 \times $$ $$( \B N \t M_+ + \A
N \t M_- )^2  (   \B^\dagger N^\dagger \t M_+^\dagger  +  \A^\dagger N^\dagger
\t M_-^\dagger )^2.
$$
Ignoring the overall sign, we obtain a term $ 16\cdot 8! \, ( | \A |^2 + | \B
|^2 )^4$ multiplied by the `volume form' for the complex fermions.  After
integrating out the fermions, the final integral reduces to,
$$ \int {8 \pi^2 \over (2\pi )^{9/2} }  ( | \A |^2 + | \B |^2 ) e^{- {1\over 2}
 ( | \A |^2 + | \B |^2 )^2}.$$
After expressing the complex coordinate $ Q$ in terms of real coordinates using
$ Q = {1\over \sqrt{2}} (Q^r + i Q^i )$, we can easily evaluate the integral
which gives one for the contribution of the principal term to the index,  again
neglecting the overall sign.

There are two issues that remain to be addressed. The first is whether the
second term in \indbound\ gives a non-vanishing contribution to the index.  One
may integrate by parts
 to show that this second integral  is proportional to

$$\lim_{R\rightarrow\infty}\int_{|x| = R}dx
\int_{U(1)} d\theta \int_0^\infty d\b \, { x^i \over R} \, \tr \,  \psi^i
(-1)^F  Q e^{-\b H} \Pi (g) \, (gx,x),
$$
where $ \psi^i$ is a fermionic operator, and $ x$ denotes all bosonic
variables.  For a more detailed discussion, see  for example, \rCallias.
Therefore, we only need to consider the kernel, $ e^{-\b H}$, at large $ R$
where $ R = |x|$.  Away from the flat points as $ R \rightarrow\infty$, the
potential term, $ e^{-\b V}$, strongly suppresses any boundary contribution.
Let us consider the theory in the neighborhood of a flat point.  Without the
mass perturbation, the bosonic potential takes the form $V \sim - {1\over 2}
r^2 |Q|^2 + O(|Q|^4) $, where $ Q$ parametrizes the transverse directions, and
$ r$ is a radial coordinate for the flat directions.  The Hamiltonian is then
essentially a set of bosonic and fermionic harmonic oscillators for the
transverse directions, and a free Laplacian along the flat directions. The two
systems are coupled through the frequency of the harmonic oscillators which
depends on the radial coordinate.  For very large $ r$, to obtain a finite
energy solution, the wavefunction in the transverse directions is approximately
the harmonic oscillator ground state, and decays very rapidly.  Clearly, the
only possible place for a boundary term to arise is from a small neighborhood
of the flat points. There are a number of arguments that suggest that there is
no contribution from around these points. For instance, after performing the
$\b$ integration, we obtain

$$\lim_{R\rightarrow\infty}\int_{|x| = R}dx
\int_{U(1)} d\theta  \,{ x^i \over R} \, \tr \,  \psi^i (-1)^F  Q H^{-1} \Pi
(g) \, (gx,x).
$$
Here $H^{-1}$ is the {\sl unbounded} operator defined to be zero on the kernel
of H and to have image orthogonal to the kernel of H.  In order to
show that this boundary term is zero, we construct an approximation G
to $H^{-1}$ with $E:= I - HG$ decaying polynomially at infinity and show that
 $$\lim_{R\rightarrow\infty}\int_{|x| = R}dx
\int_{U(1)} d\theta  \, { x^i \over R} \, \tr \,  \psi^i (-1)^F  Q G \Pi (g) \,
(gx,x) = 0.
$$
 The error in this aproximation is then
 $$\lim_{R\rightarrow\infty}\int_{|x| = R}dx
\int_{U(1)} d\theta  \, { x^i \over R} \, \tr \,  \psi^i (-1)^F  Q H^{-1}E \Pi
(g) \, (gx,x),
$$
which one expects to vanish (but we shall not establish this rigorously at this
 time).
The operator G is obtained in a standard iterative construction, except that
rather than Fourier expanding in all variables we Fourier expand in the $x,y$
directions and use the natural harmonic oscillator expansion in the $ Q $
directions.  Following the approach outlined above, we have found no correction
to the index for this model.

So far, we have counted ground states in an index sense. We would like to argue
that the index is actually counting the total number of ground states.
However, there does not seem to be a simple vanishing theorem to guarantee that
there are no ground states which are either fermionic or bosonic in this
particular  model.  Agreement with duality suggests that the ground state that
we have found is unique. Nevertheless, we can conclude that there is at least
one normalizable ground state from each of the sixteen singularities. These
modes provide the missing H-monopole states as required by the conjectured
S-duality of the toroidally compactified heterotic string.

\vskip 0.2in\noindent
{\bf Note added:}
While we were completing this work, a paper \rMP\  appeared which discusses
this problem from a somewhat different viewpoint.

\bigbreak\bigskip\bigskip\centerline{{\bf Acknowledgements}}\nobreak
We would like to thank A. Lesniewski, S. Mathur and E. Witten for helpful
discussions and comments.  The work of S. S. was supported in part by a Hertz
Fellowship and NSF grant PHY-92-18167; that of M. S. by NSF Grant DMS 9505040.

\listrefs

\end